\documentclass[12pt,thmsa]{article}

\input{tcilatex}

\begin{document}

\begin{center}
{\LARGE Chirality and helicity in terms of}

{\LARGE \ topological spin and topological torsion}\vspace{1pt}\bigskip

\textbf{R. M. Kiehn}

University of Houston

rkiehn2352@aol.com

\vspace{1pt}
\end{center}

\begin{quote}
\textbf{Abstract:} In this article the concept of enantiomorphism is
developed in terms of topological, rather than geometrical, concepts. \
Chirality is to be associated with enantiomorphic pairs which induce Optical
Activity, while Helicity is to be associated enantiomorphic pairs which
induce a Faraday effect. Experimentally, the existence of enantiomorphic
pairs is associated with the lack of a center of symmetry, which is also
serves as a necessary condition for Optical Activity. However, Faraday
effects may or may not require a lack of a center of symmetry. The two
species of enantiomorphic pairs are distinct, as the rotation of the plane
of polarization by Optical Activity is a reciprocal phenomenon, while
rotation of the plane of polarization by the Faraday effect is a
non-reciprocal phenomenon. \ From a topological viewpoint, Maxwell's
electrodynamics indicates that the concept of Chirality is to be associated
with a third rank tensor density of Topological Spin induced by the
interaction of the 4 vector potentials \{$\mathbf{A}$, $\phi $ \} and the
field excitations ($\mathbf{D,H}$). The distinct concept of Helicity is to
be associated with the third rank tensor field of Topological Torsion
induced by the interaction of the 4 vector potentials and field intensities (%
$\mathbf{E,B}$).

\bigskip
\end{quote}

\section{Introduction}

\vspace{1pt}It is a remarkable result of experimental chemistry, recognized
by Pasteur and others, that there can exist enantiomorphic pairs of states
of chemical systems that cannot be smoothly mapped into one another,
starting from the identity. \ From a geometrical perspective, right handed
quartz and left handed quartz are systems with apparent equivalent energies,
yet with decidedly different behavior when interacting with electromagnetic
fields. \ If fact, the rotation of the plane of optical polarization is
often used as a tool to distinguish the enantiomorphic isotopes. \ However,
the issue is deeper than that of geometry. \ There exist topologically
equivalent isotopes that cannot be smoothly mapped into one another starting
from the identity. \ Geometrical properties are those subsets of topological
properties that depend upon size and shape. However, herein, the category of
interest is that subset of topological properties which do not depend upon
the geometrical issues of size and shape, and yet are to be associated with
enantiomorphic pairs. \ These distinct topological isotopes (enantiomers)
are of different size and shape and are connected by homeomorphisms, but
they are not smoothly connected to one another by a map about the identity.
\ The properties and existence of such topological enantiomers is the theme
of this article. \ For example, a right handed Moebius band is topologically
equivalent to a left handed Moebius band, but possibly of different size and
distorted shape. \ The homeomorphism between the two topological isotopes
consists of more than one step: the right handed Moebius band can be
transformed into a left handed Moebius band by first cutting the band,
applying a 360 degree twist, and then reconnecting the cut ends such that
points that were initially near to one another remain near to one another. \
Such a combination of processes is not C2 smooth but can be continuous in
the topological sense.

The objective of this article is to examine those special features of
electromagnetic systems that can exhibit topological enantiomers, and to
determine how such enantiomers can be created. \ As all chemical systems are
special examples of electromagnetic interactions, the methods to be
developed are useful to the understanding of the more constrained
(geometrical)\ features of chemical enantiomorphism. \ A remarkable, but
little appreciated, fact is that the electromagnetic field itself has
certain properties that exhibit the topological enantiomorphism mentioned
above. \ Hence a study of these electromagnetic properties (defined below as
topological torsion and topological spin) delivers a necessary foundation
for the existence, control and modification of the geometric properties of\
enantiomorphism displayed in modern chemistry.

In modern steriochemistry, Optical Activity and Faraday Rotation have a
dominant experimental role. \ Optical Activity and Faraday Rotation have
many similarities, yet they are distinct, different, electromagnetic
phenomena. \ A necessary, but not sufficient, condition for Optical Activity
in crystalline structures is the lack of a center of symmetry. \ This lack
of a center of symmetry is often used as the basis for defining
''chirality'', and, conversely, chirality is often associated with Optical
Activity. \ \ However, according to Post \lbrack 1\rbrack , there exist 3
crystal classes without a center of symmetry (crystal classes 26,27,29) that
do not support Optical Activity, hence a lack of a center of symmetry is not
sufficient condition for Optical Activity. \ Note that Optically Active
media have the capability of ''rotating'' the plane of polarization, as
linearly polarized light passes through the media, a practical effect used
by the wine grower to estimate the sugar content in his grapes.

Similar rotation of the plane of polarization occurs when linearly polarized
light passes through Faraday media. \ However, Faraday effects can exist
both in crystalline structures that have a center of symmetry and in
crystalline structures that do not have a center of symmetry. \ Post reports
that there are 9 crystalline structures that support both optical Activity
and Faraday rotation. \ What are the intrinsic differences between Optical
Activity and Faraday Rotation? \ \vspace{1pt}In his book ''Formal Structure
of Electromagnetics'', E. J. Post \lbrack 1\rbrack\ clearly delineates the
differences between Optical Activity and Faraday rotation, and demonstrates
solutions to Maxwell's equations for both effects. \ The crucial result is
that Optical Activity is reciprocal and Faraday Rotation is not.

In short, the lack of a center of symmetry, the rotation of the plane of
polarization, and the existence of enantiomorphic pairs, are necessary but
not sufficient properties to define the concept of chirality. \ There exist
two species of phenomena that exhibit the three properties stated above, one
species is ''reciprocal'' and defines Chirality, and the other species is
''non-reciprocal'' and defines Helicity. \ These differences between
Chirality and Helicity deserve attention, clarification, and exploitation. \
Such is the purpose of this article. \ 

\subsection{\qquad Transverse Inbound and Outbound Waves}

First consider a complex four vector potential solution to the vector wave
equation which propagates as a transverse wave in the $\pm z$ direction with
a phase $\theta =\pm kz\mp \omega t.\,$\ There are 4 possibilities: \ The $%
\mathbf{E}$ field rotates about the $z$ axis in a Right Handed manner as
viewed by an observer looking towards the positive $z$ direction, or it
rotates in a Left Handed manner. \ \ Outbound $\theta =kz-\omega t$ and
Inbound $\theta =-kz-\omega t\,$\ waves are to be distinguished as ORH, OLH,
IRH, and ILH..

\begin{eqnarray}
ORH &=&\left| 
\begin{array}{l}
1 \\ 
i
\end{array}
\right\rangle e^{i(kz-\omega t)}\,\,\,\,\,\,\,\,\,\ IRH=\left| 
\begin{array}{l}
1 \\ 
i
\end{array}
\right\rangle e^{i(-kz-\omega t)}\,\,\,\,\,\,\, \\
\,\,\,\,\,OLH &=&\left| 
\begin{array}{l}
1 \\ 
-i
\end{array}
\right\rangle e^{i(kz-\omega t)}\,\,\,\,\,\,\,I\,LH=\left| 
\begin{array}{l}
1 \\ 
-i
\end{array}
\right\rangle e^{i(-kz-\omega t)}
\end{eqnarray}
For media with the symmetries of the Lorentz vacuum, the phase velocities $%
v=\omega /k$ are the same for all four modes. \ Addition or subtraction of
ORH and OLH produces a Linearly polarized state outbound. \ Addition or
subtraction of IRH and ILH produces a Linearly polarized state inbound. \ 

Next, recall the experimental differences between Optical Activity and
Faraday Rotation:

\subsubsection{\qquad Optical Activity}

Consider an optically active fluid (sugar in water) in a cylindrical tube of
length L. \ For Optical Activity, there are also two distinct phase
velocities, $\omega /k_{1}$ and $\omega /k_{2}.$\ Outbound Right Handed
(ORH) circularly polarized light propagates with a phase speed equal to the
phase speed of Inbound Left Handed (ILH) circularly polarized light. \
Outbound Left Handed (OLH) polarized light propagates with a phase velocity
different from the phase velocity of Outbound Right Handed polarized light
(ORH), but with the same speed as that of Inbound Right Handed (IRH)
polarized light. \ In summary,

\begin{equation}
\text{Optical Activity Phase Velocity, \ \ }V_{ORH}=V_{ILH}\,\neq
V_{OLH}\,=V_{IRH}\,\,\,\,
\end{equation}

The wave solutions for optical activity are of the format:

\begin{eqnarray}
ORH &=&\left| 
\begin{array}{l}
1 \\ 
i
\end{array}
\right\rangle exp\,i(k_{1}z-\omega t)\,\,\,\,\,\,\,\ IRH=\left| 
\begin{array}{l}
1 \\ 
i
\end{array}
\right\rangle exp\,i(-k_{2}z-\omega t)\,\,\,\,\,\,\,\,\,\,\,\, \\
OLH &=&\left| 
\begin{array}{l}
1 \\ 
-i
\end{array}
\right\rangle exp\,i(k_{2}z-\omega t)\,\,\,\,\,\,\,\,ILH=\left| 
\begin{array}{l}
1 \\ 
-i
\end{array}
\right\rangle exp\,i(-k_{1}z-\omega t)  \nonumber
\end{eqnarray}

\vspace{1pt}The addition of $\left| ORH\right\rangle +\left|
OLH\right\rangle $ produces a Linearly Polarized state propagating outbound,
whose plane of polarization rotates. \ When the two inbound states are
added, $\left| IRH\right\rangle +\left| ILH\right\rangle ,$ a linearly
polarized state is achieved, and its plane of polarization also rotates 
\textit{but in the opposite direction as the outbound rotation. \ }In other
words, the round trip (outbound+reflection+inbound) motion causes the plane
of polarization to return to its initial value. \ This result defines what
is meant by a reciprocal effect. If the plane of polarization of the
original linearly polarized light beam suffers a rotation in the amount of $%
\theta $ degrees as it traverses the Optically Active media, when reflected
in a mirror, the plane of polarization suffers an negative rotation of $%
\theta $ degrees, as the light beam traverses the media in the reverse
direction. \ The plane of polarization returns to its original state after
the round trip. \ (The sense of Right Handed and Left Handed polarization is
determined by an observer looking away from himself.)$\,\,\,\,\,\,\,\,$

\subsection{\qquad Faraday Rotation}

\vspace{1pt}\ Consider a gas of He-Ne in a cylindrical tube of length L. \
Surround the tube with a coil of wire that will produce a coaxial magnetic
field that partially aligns the spins of the gas atoms. \ \vspace{1pt}For
such Faraday media, there are two distinct phase velocities, $\omega /k_{1}$
and $\omega /k_{2}$. \ Outbound Right Handed (ORH) circularly polarized
light propagates with a phase speed equal to the phase speed of Inbound
Right Handed (IRH) circularly polarized light. \ Outbound Left Handed (OLH)
polarized light propagates with a phase velocity different from the phase
velocity of Outbound Right Handed polarized light (ORH), but with the same
speed as that of Inbound Left Handed (ILH) polarized light. \ \ In summary,

\begin{equation}
\text{Faraday Effect Phase Velocity, \ \ }V_{ORH}=V_{IRH}\,\neq
V_{OLH}\,=V_{ILH}\,\,\,\,
\end{equation}

\vspace{1pt} The wave solutions for the Faraday effect are of the format:

\begin{eqnarray}
\left| ORH\right\rangle &=&\left| 
\begin{array}{l}
1 \\ 
i
\end{array}
\right\rangle exp\,i(k_{1}z-\omega t)\,\,\,\,\,\,\,\ \left| IRH\right\rangle
=\left| 
\begin{array}{l}
1 \\ 
i
\end{array}
\right\rangle exp\,i(-k_{1}z-\omega t)\,\,\,\,\,\,\,\,\,\,\,\, \\
\left| OLH\right\rangle &=&\left| 
\begin{array}{l}
1 \\ 
-i
\end{array}
\right\rangle exp\,i(k_{2}z-\omega t)\,\,\,\,\,\,\,\,\left| ILH\right\rangle
=\left| 
\begin{array}{l}
1 \\ 
-i
\end{array}
\right\rangle exp\,i(-k_{2}z-\omega t)  \nonumber
\end{eqnarray}
\ The formulas represent circularly polarized waves. The addition of $\left|
RHO\right\rangle +\left| LHO\right\rangle $ produces a Linearly Polarized
state propagating outbound, whose plane of polarization rotates. \ When the
two inbound states are added, $\left| RHI\right\rangle +\left|
LHI\right\rangle ,$ a linearly polarized state is achieved, and its plane of
polarization also rotates \textit{in the same direction as the outbound
rotation. \ }In other words, the round trip (outbound+reflection+inbound)
motion does not cause the plane of polarization to return to its initial
value. \ This result defines what is meant by a non-reciprocal effect. \ If
the plane of polarization of the original linearly polarized light beam
suffers a rotation in the amount of $\theta $ degrees as it traverses the
Faraday media, when reflected in a mirror, the plane of polarization suffers
an additional rotation of $\theta $ degrees, as the light beam traverses the
media in the reverse direction. \ The plane of polarization does not return
to its original state, but instead ratchets by $2\theta $ degrees upon
completing the round trip. \ (The sense of Right Handed and Left Handed
polarization is determined by an observer looking away from himself.)$%
\,\,\,\,\,\,\,\,$

\subsection{\protect\vspace{1pt}Polar and Axial vectors}

\vspace{1pt}Following Schouten \lbrack 2\rbrack , Post points out that
Faraday Rotation is ''generated'' by a ''W vector'', while Optical Activity
is generated by a ''vector''. \ Under certain constraints, the W\ vector
plays the role of an ''Axial'' vector, while the ''vector'' becomes a
''polar'' vector. \ Upon reflection, a polar vector changes its sense
(determined by the arrow head). \ Point your finger into a mirror. The image
points back at you. \ The sense of the image is opposite to the sense of the
object. \ For polar vectors with a line of action parallel to the mirror
surface, the opposite result is obtained. \ The sense of the image is the
same as the sense of the object. \ Note the differences of orthogonal and
parallel reflections.

A reflected axial vector does not change its sense if the line of action is
orthogonal to the mirror. \ Curl you fingers and align your thumb in a
direction orthogonal to the mirror. \ It does not matter whether the thumb
points into or away from the mirror. \ The sense of the ''axial vector'' is
determined by the curl of the fingers. The sense of the reflected image is
the same as the sense of the object. \ The opposite effect occurs when the
line of action of the axial vector is parallel to the reflection surface. \
The sense, as determined by the curl of the fingers, is opposite to that of
the reflected image.

The magnetic field $\mathbf{B}$ and the angular velocity $\mathbf{\Omega \,\,%
}$are examples of spatial ''W vectors''. \ On the other hand, the $\mathbf{D}
$ field is a spatial ''polar vector'' in the sense used by Post. \ The
anti-symmetric spatial components of the covariant field intensity tensor
2-form, $F=dA,$ are formed by the spatial ''W vector'' field $\mathbf{B.}$ \
The anti-symmetric spatial components of the tensor density, N-2 form, $G,$
where $J=dG,$ are formed by the spatial ''polar vector'' field $\mathbf{D.}$
\ These facts yield a clue for distinguishing Faraday Rotation and Optical
Activity on topological grounds. \ As will be shown below, Faraday Rotation
is to be associated with the concept of Topological Torsion, and Optical
Activity is to be associated with the concept of \ Topological Spin.

\section{Topological Formulation of Maxwell's Equations.}

\subsection{ \ \ \ \ \ Exterior Differential Systems}

It is known that Maxwell's system of PDE's (without constitutive
constraints) can be expressed as an exterior differential system \lbrack
3\rbrack\ on a variety of independent variables. \ Exterior differential
systems impose topological constraints on a differential variety. \ For the
Maxwell electromagnetic system on a domain \{x,y,z,t\} the two topological
constraints have been called the Postulate of Potentials, and the Postulate
of Conserved Currents. \lbrack 4\rbrack . \ These two topological
constraints lead to the system of Partial Differential Equations, known as
Maxwell's equations, for any coordinate system so constrained. \ No metric,
no connection, nor other restraints of a geometrical nature are required on
the 4 dimensional differential variety of independent variables, typically
written as $\{x,y,z,t\}.$

\begin{equation}
\text{ Postulate of Potentials (an exact 2-form)\ \ \ \ \ \ \ \ \ \ \ \ \ \
\ \ \ \ \ \ \ }F-dA=0
\end{equation}

\begin{equation}
\text{ Postulate of Conserved Currents (an exact 3-form) \ \ \ \ \ }J-dG=0
\end{equation}

\ \ The method of exterior differential systems insures that the description
is not\ only diffeomorphically invariant in form (natural covariance of form
with respect to all invertible smooth coordinate transformations), but also
the description is functionally well defined with respect to maps which are
C2 continuous, but not necessarily invertible. \ This statement implies that
those exterior differential forms which are defined on a final state variety
can be ''pulled back'' in a functionally well defined manner to an initial
state variety, even though the map from initial to final state of coordinate
variables is NOT a diffeomorphic coordinate transformation. \ The inverse
mapping need not exist. \ This result is truly a remarkable property of
Maxwell electrodynamics, for it permits the analysis of certain irreversible
electrodynamic processes without\ the use of statistics. \ The ''push
forward'' process is not functionally well defined when the inverse map does
not exist, a fact that demonstrates that topological evolution induces an
''arrow of time'' \lbrack 5\rbrack .

\subsection{ \ \ \ \ \ Constitutive Constraints}

In practical applications, it is possible to impose constraints on the
Maxwell system in the form of constitutive relations between the
thermodynamically conjugate variables of field intensity $\mathbf{(E,B)}$
and field excitations $\mathbf{(D,H)}$. \ \ Post has demonstrated that the
constitutive tensor (density) has many of the properties of the Riemann
tensor \lbrack 6\rbrack . \ \ \ These constraints are NOT necessarily
equivalent to the Riemann tensor generated by a Riemannian metric imposed
upon the variety $\{x,y,z,t\}.$ \ \ In many circumstances the equivalence
classes of such constitutive constraints can\ be put into correspondence
with the geometrical symmetries of the 32 crystal classes that are used to
discriminate between the many different observed physical structures. \ As
mentioned above, a \textit{complex} 6x6 constitutive constraint has been
used by Post to delineate between Optical Activity, Faraday Phenomena,
Birefringence and Fresnel-Fizeau motion induced effects in electromagnetic
signal propagation. \ The complex constitutive tensor cannot be deduced from
a real metric tensor. \ However it would appear that the constitutive tensor
has a constructive definition in terms of a non-symmetric connection.

Indeed, the work of Post, who subsumed a complex constitutive tensor, has
been extended \lbrack 7\rbrack\ to demonstrate the existence of \
irreducible ''quaternion'' solutions to the Maxwell system. \ Quaternion
waves cannot be represented by complex functions, which are the usual choice
for describing electromagnetic signals. \ Complex wave solutions generate a
4th order characteristic polynomial for the phase speed which is doubly
degenerate. \ The wave speeds have only two distinct magnitudes depending
upon direction and polarization. \ For cases where a center of symmetry is
not available, and yet the medium supports both Optical Activity and Faraday
rotation, the wave solutions can NOT\ be expressed as complex functions, but
can be written as quaternions. \ The resulting 4th order characteristic
polynomial for the wave speeds is not degenerate and has four distinct root
magnitudes. \ The results indicate that the phase propagation speed of light
is different for each direction of propagation and for each mode of
polarization. \ The theory has been used to explain the experimental results
measured in dual polarized ring laser apparatus.

In contrast, in a medium with the Lorentz symmetries, the characteristic
polynomial is 4-fold degenerate; e.g., all polarizations and all \
directions have the same propagation speed. \ The result leads to the
ubiquitous statement that the speed of light is the same for all observers,
which is incorrect for media that do not have the Lorentz symmetries. \ \
For Birefringent, or Optically Active, or Faraday media, the characteristic
polynomial for phase velocity is doubly degenerate, implying a relationship
exists between for the 4 modes of propagation. \ There exist only two
distinct phase velocity magnitudes. \ \ The correlation speeds for direction
and polarization pairs have been presented above. \ Faraday rotation and
Optical Activity have different propagation direction-polarization
handedness correlations. \ The Faraday rotation is not reciprocal; \ the
rotation induced by Optical Activity is reciprocal.

\subsection{Topological \protect\vspace{1pt}Three Forms}

\vspace{1pt}The classic formalism of electromagnetism is a consequence of a
system of two fundamental topological constraints as defined above on a
domain of four independent variables. The theory requires the existence of
two fundamental exterior differential forms, $\{A,G\},$ where the postulates
permit the construction of the differential ideal $\{A,F=dA,G,J=dG\}.$ \
This system of differential forms may be prolonged (by construction all
possible exterior products) to yield the Pfaff sequence of forms: $%
\{A,F,G,J,A\symbol{94}G,A\symbol{94}F,A\symbol{94}G,F\symbol{94}F,F\symbol{94%
}G,A\symbol{94}J,G\symbol{94}G\}$. \ On a domain of four independent
variables, the complete Pfaff sequence contains three 3-forms: the classic
3-form of charge current density, $J,$ and the (apparently novel to many
researchers) \ 3-forms of Spin Current density, $A\symbol{94}G,$[9]\ and
Topological Torsion-Helicity, $A\symbol{94}F$\ \ [10]. \ 
\begin{equation}
\text{The 3-form of Charge Current density \ \ \ }J=dG
\end{equation}
\begin{equation}
\text{The 3-form of Topological Spin density \ \ \ }S=A\symbol{94}G
\end{equation}
\begin{equation}
\text{The 3-form of Topological Torsion \ \ \ \ \ \ \ }T=A\symbol{94}F
\end{equation}
In most elementary descriptions of electromagnetic theory, the 3-forms of
Spin and Torsion are ignored. \ \ By direct evaluation of the exterior
product, and on a domain of 4 independent variables, each 3-form will have 4
components that can be symbolized (in engineering format) by the 4-vector
arrays

\begin{equation}
Spin-Current:\mathbf{S}_{4}=[\mathbf{A\times H}+\mathbf{D}\phi ,\mathbf{%
A\circ D}]\equiv \lbrack \mathbf{S,}\sigma \mathbf{]},
\end{equation}
\vspace{1pt}and

\begin{equation}
Torsion-vector:\mathbf{T}_{4}=[\mathbf{E\times A}+\mathbf{B}\phi ,\mathbf{%
A\circ B}]\equiv \lbrack \mathbf{T,}h\mathbf{]},
\end{equation}
which are to be compared with the four construction components of the charge
current 4-vector density:

\begin{equation}
Ch\arg e-Current:\mathbf{J}_{4}=[\mathbf{J},\rho ].
\end{equation}

\subsection{ \ \ \ \ \ Topological Invariants}

The closed integral of each of the three 3-forms is a deformation invariant
(hence a topological property) if the selected 3-form is closed in an
exterior derivative sense ($dJ=0,\,\,dS=0,\,\,dT=0\,\ $respectively). \ \
For example, for any 3-form, $J,$ such that $dJ=0,$ the Lie derivative of
the closed integral relative to an arbitrary process path denoted by $\beta
V $ is given by the expression,

\begin{equation}
\vspace{1pt}L_{\beta \mathbf{V}}\tiiint_{closed}\,(J)=\tiiint_{closed}\{i(%
\beta V)d(J)+d(i(\beta V)(J)\}=\tiiint_{closed}\{0+d(i(\beta V)(A\symbol{94}%
G)\}=0.
\end{equation}
The zero result is interpreted by the statement ''the closed integral is a
deformation invariant'' for the process can be deformed by any non-zero
function $\beta (x,y,z,t),$ and the integral is unchanged.\vspace{1pt}

As the charge current 3-form, $J,$ is closed by construction, $(dJ=ddG=0)$
it follows that its closed integral is always a deformation invariant. \ The
result leads to another ubiquitous statement known in electromagnetic theory
as the ''Conservation of electric charge''. \ It is not equivalent to the
quantization of charge. \ The additional topological constraints of closure
imply that the exterior derivative of each of the three forms is empty
(zero). \ \ By direct computation, such a constraint of differential closure
leads to the Poincare invariants for the electromagnetic system.

\begin{eqnarray}
Poincare\,\,1 &=&d(A\symbol{94}G)=F\symbol{94}G-A\symbol{94}J  \nonumber \\
&=&\{div_{3}(\mathbf{A\times H}+\mathbf{D}\phi )+\partial (\mathbf{A\circ D)}%
/\partial t\}dx\symbol{94}dy\symbol{94}dz\symbol{94}dt  \nonumber \\
&=&\{(\mathbf{B\circ H-D\circ E)-(A\circ J}-\rho \phi )\}dx\symbol{94}dy%
\symbol{94}dz\symbol{94}dt
\end{eqnarray}

\begin{eqnarray}
Poincare\,\,2 &=&d(A\symbol{94}F)=F\symbol{94}F  \nonumber \\
&=&\{div_{3}(\mathbf{E\times A}+\mathbf{B}\phi )+\partial (\mathbf{A\circ B)}%
/\partial t\}dx\symbol{94}dy\symbol{94}dz\symbol{94}dt  \nonumber \\
&=&\{-2\mathbf{E\circ B}\}dx\symbol{94}dy\symbol{94}dz\symbol{94}dt
\end{eqnarray}

For a (vacuum) state, with $J=0,$ zero values of the Poincare invariants
require that the magnetic energy density is equal to the electric energy
density $(1/2\mathbf{B\circ H}=1/2\mathbf{D\circ E)}$, and, respectively,
that the electric field is orthogonal to the magnetic field $(\mathbf{E\circ
B}=0\mathbf{).}$ \ Note that these constraints often are used as elementary
textbook definitions of \ what is meant by electromagnetic waves. \ The
possible values of the topological quantities, as deRham period integrals
\lbrack 11\rbrack , form rational ratios and topological quantum numbers. \
These quantum numbers should NOT\ be considered as topological ''charge''. \
Electromagnetic (topological) charge is related to the two dimensional
closed integrals of $G,\,$not the three dimensional closed integrals
described above.

\subsection{\ \ Field Momentum, Propagation direction, and the 4-Vector
Potential}

The 4 vector potential $A_{4}=[\mathbf{A},\phi ]$\ is different from the 4
dimensional propagation vector $k_{4}=[\mathbf{k},\omega ];$ \ \ in many
cases the two vectors are not even proportional (although they can be). \ In
the language of differential forms, it must be recognized that there is a
difference between the 1-form, $A=A_{x}dx+A_{y}dy+A_{z}dz-\phi dt,$ and the
1-form, $k=k_{x}dx+k_{y}dy+k_{z}dz-\omega dt.$ \ The integral of $k$ defines
the phase $\theta =\int k.$ \ \ The propagation 1-form $k$ is defined from
the equations that generate the singular solutions [7]\ to Maxwell's
equations: 
\begin{equation}
k\symbol{94}F=0,\text{ \ \ and \ \ }k\symbol{94}G=0.
\end{equation}
Note that these equations for singular solutions are derived from the
3-forms\ $k\symbol{94}F$ and $k\symbol{94}G$ constrained to be zero. \ The
3-form $k\symbol{94}F$ \ is not necessarily the same as the 3-form, $A%
\symbol{94}F.$ \ \ Similarly, the 3-form $k\symbol{94}G$ \ is not
necessarily the same as the 3-form, $A\symbol{94}G.$ \ \ These singular
solutions are always transverse in a geometrical sense that the wave 3
vector $\mathbf{k}$ is in the direction of the field momentum, $\mathbf{D}%
\times \mathbf{B}.$ \ Both the $\mathbf{D}$ vector and the $\mathbf{B}$
vector are orthogonal to the wave vector, $\mathbf{k}$, that generates
singular solutions. \ 

The 1-forms, $k$, that satisfy the equations 
\begin{equation}
\text{Associated 1-forms k }\{k\symbol{94}F=0,k\symbol{94}G=0\}
\end{equation}
are defined as the ''associated'' 1-forms relative to$\,F$\ and $G$. \ The
1-forms, $k$, that satisfy the equations 
\begin{equation}
\text{ Extremal 1-forms k \ }\{k\symbol{94}dF=0,k\symbol{94}dG=0\}
\end{equation}
are defined as ''extremal'' 1-forms relative to $F$\ and $G$. \ 1-forms such
that 
\begin{equation}
\text{characteristic 1-forms k \ \ }\{k\symbol{94}F=0,k\symbol{94}G=0,\,k%
\symbol{94}dF=0,\;k\symbol{94}dG=0\},
\end{equation}
are defined as the ''characteristic'' 1-forms, $k.$ \ These 1-forms $k$ are
dual to the associated, extremal, and characteristic vector fields, $V$,
which satisfy the equations 
\begin{equation}
\text{ Associated vector fields V \ \ \ }\{i(V)A=0,\,i(V)G=0\}
\end{equation}
or 
\begin{equation}
\text{Extremal vector fields V \ \ \ \ }\{i(V)dA=0,\,i(V)dG=0\},
\end{equation}
or 
\begin{eqnarray}
\text{ Characteristic vector fields V for A\ \ \ \ }\{i(V)A &=&0,\,\ \ \ \ \
i(V)dA=0, \\
\text{Characteristic vector fields V for G \ \ \ \ }i(V)G &=&0,\,\ \
\;i(V)dG=0\},
\end{eqnarray}
respectively.

The concepts of \ Spin Current and the Torsion vector have been utilized
hardly at all in applications of classical electromagnetic theory. \ Just as
the vanishing of the 3-form of charge current, $J=0,$ defines the
topological domain called the vacuum, the vanishing of the two other 3-forms
will refine the fundamental topology of the Maxwell system.\ \ \ Such
constraints permit a definition of transversality to be made on topological
(rather than geometrical) grounds. If both $A\symbol{94}G$ and $A\symbol{94}%
F $ vanish, the vacuum state supports topologically transverse modes only
(TTEM). \ Examples lead to the conjecture that TTEM modes do not transmit
power, a conjecture that has been verified when the concept of geometric
transversality (TEM) and topological transversality (TTEM) coincide. \ \ A
topologically transverse magnetic (TTM) mode corresponds to the topological
constraint that $A\symbol{94}F=0.$ \ A topologically transverse electric
mode (TTE) corresponds to the topological constraint that $A\symbol{94}G=0.$%
\ \ Examples, both novel and well-known, of vacuum solutions to the
electromagnetic system which satisfy (and which do not satisfy) these
topological constraints are given [12]. \ The ideas should be of interest to
those working in the field of Fiber Optics. \ Recall that classic waveguide
solutions which are geometrically and topologically transverse (TEM$\equiv $%
TTEM) do not transmit power [13]. \ \ However, in [12]\ an example vacuum
wave solution is given which is geometrically transverse (the fields are
orthogonal to the field momentum and the wave vector), and yet the
geometrically transverse wave transmits power at a constant rate: \ the\
example wave is not topologically transverse as $A\symbol{94}F\neq 0.$ \ 

\subsection{\protect\vspace{1pt}Connections for Right handed vs Left handed
evolution}

In spaces (such as Finsler spaces) which may or may not be Riemannian, the
topological concept of differential neighborhoods that are linearly
connected implies the existence, over the domain, of a matrix of functions $%
\left[ F_{a}^{k}(q^{b})\right] $ that will linearly map 1-forms (linear
combinations of differentials $\left| \omega ^{a}(q,dq)\right\rangle $) into
1-forms (linear combinations of differentials $\left| \sigma
^{k}(q,dq)\right\rangle $). \ \vspace{1pt} 
\begin{equation}
\left[ F_{a}^{k}(q^{b})\right] \circ \left| \omega ^{a}(q,dq)\right\rangle
\Rightarrow \left| \sigma ^{k}(q,dq)\right\rangle .
\end{equation}
The map between differentials is linear, but the matrix elements are not
domain constants (non-linearity is built in). \ The columns of the matrix of
functions forms a matrix of basis vectors over the domain which may be used
to express any tensorial properties. \ The matrix of basis vectors defines
what Cartan called the Repere Mobile, or the moving frame, for its values
change as a point p moves along some curve in the domain. \ If the mapping
of 1-forms is integrable,

\begin{equation}
integrable\,basis:\,\,\,\,\,\,\,\,\left[ F_{a}^{k}(q^{b})\right] \circ
\left| dq^{a}\right\rangle \Rightarrow \left| dx^{k}\right\rangle
,\,\,\,\,\,implies\,\,\,\ x^{k}=f^{k}(q^{b})
\end{equation}
then there exist functions whose differentials are exact, and the mapping is
said to be holonomic. \ Often it is presumed that the functional mapping
exists (and as such is called a coordinate transformation under certain
additional constraints) and the Frame matrix is deduced by the differential
operations to produce the Jacobian matrix of the transformation. \ However,
there are other ways to impose or deduce a Frame matrix on a domain.

The fundamental question of a connection is related to the differential
neighborhood properties of the Frame matrix. \ As the Frame matrix, by
definition of its domain, has a non-zero determinant, then it admits an
inverse matrix, and by differential and algebraic processes the (right
Cartan)\ connection matrix $\left[ C_{R}\right] $ (or the left Cartan matrix 
$\left[ C_{L}\right] )$ can be constructed:

\begin{eqnarray}
d\left[ F\right] &=&-\left[ F\right] \circ \left[ dG\right] \circ \left[ F%
\right] \\
&=&-\left[ C_{L}\right] \circ \left[ F\right]  \nonumber \\
&=&+\left[ F\right] \circ \left[ C_{R}\right] .  \nonumber
\end{eqnarray}
The matrix elements of $\left[ C_{R}\right] $ and $\left[ C_{L}\right] $ are
differential 1-forms. \ These matrix elements do not necessarily vanish nor
are they necessarily equal. \ It is to be noted that the left and the right
Cartan matrices are (anti-) similarity transforms of one another. \ 

\vspace{1pt}The reason that there are two distinct methods for constructing
the linear mapping is based upon the fact that a matrix, whose determinant
is non - zero, always has two representations, a left handed and a right
handed representation. \ The two representations always consist of either a
(right handed) product of a unitary matrix times a Hermitean matrix, or the
(left handed)\ product of a Hermitean matrix times a unitary matrix. Only if
the original matrix is ''normal'' ( such that the product of itself times
its Hermitian conjugate is equal to the product of its Hermitian conjugate
times itself) will the right and left handed product representations be
degenerately the same [8].\ \ It is this handedness property of topological
neighborhoods that captures the features of electromagnetic charge
distributions of molecules and crystals that exhibit enantiomorphic states.
\ 

In the electromagnetic situation, the constitutive map is often considered
to be (within a factor) a linear mapping between two six dimensional vector
spaces. \ As such the constitutive map can have both a right or a left
handed representation, implying that there are two topologically equivalent
states that are not smoothly equivalent about the identity. \ 

In the geometric situation, the matrix elements of frame matrix, $\left[
F_{a}^{k}(q^{b})\right] ,$ are not constants. \ The two mechanisms (right
and left handed) for neighborhood expressions imply that the connection is
not generated from a symmetric metric. \ Such connections are said to admit
torsion. \ In short, the concept of an affine connection is more general
than the Hermitian (symmetric) connection offered by the Christoffel symbols
(\ which are generated from a metric). \ Any domain which is parallelizable
will support a linear connection of differentials. \ 

\bigskip

\section{Acknowledgments}

This work was presented at the International Symposium on Chirality, \ ISCD
12 --- \ Chirality 2000, Chamonix, France

\section{References}

\lbrack 1\rbrack\ Post, E. J., (1997), ''Formal Structure of
Electromagnetics'', Dover \ p 166.

\lbrack 2\rbrack\ Schouten, J. A., (1954), ''Tensor Analysis for
Physicists'', Dover p.32

\lbrack 3\rbrack\ \ Bryant, R.L.,Chern, S.S., Gardner, R.B.,Goldschmidt,
H.L., and Griffiths, P. A., (1991), ''Exterior Differential Systems'',
Springer Verlag.

\lbrack 4\rbrack\ \ \ Kiehn, R. M., (1998),
http://www22.pair.com/csdc/pdf/classice.pdf

\lbrack 5\rbrack\ \ \ Kiehn, R. M., (1976), ''Retrodictive Determinism'',
Int. J. of Eng. Sci. \textbf{14}, p. 749

\lbrack 6\rbrack\ \ \ Ibid, Post, p. 188

\lbrack 7\rbrack\ \ \ Kiehn, R. M., Kiehn, G. P., and Roberds, R. B. (1991)
''Parity and Time-reversal Symmetry Breaking, Singular Solutions'', Phys Rev
A,\ \textbf{43}, p. 5665

\lbrack 8\rbrack\ \ Turnbull, H. W. and Aitken, A. C., (1961) ''An
Introduction to the Theory of Canonical Matrices'', Dover, p194.

\lbrack 9\rbrack\ \ \ Kiehn, R. M. (1977) ''Periods on manfolds,
quantization and gauge'', J. of Math Phys 18, no. 4, p. 614

\lbrack 10\rbrack\ Kiehn, R. M., (1990) ''Topological Torsion, Pfaff
Dimension and Coherent Structures'', in: H. K. Moffatt and T. S. Tsinober
eds, Topological Fluid Mechanics, Cambridge University Press, 449-458 .

\lbrack 11\rbrack\ \ deRham,G. (1960) ''Varietes Differentiables'', Hermann,
Paris

[12]\ \ Kiehn, R. M., ''Torsion and Spin as topological coherent structures
in a plasma'' (1998), http://www22.pair.com/csdc/pdf/plasma.pdf

....(1999) '' Topological evolution of classical electromagnetic fields and
the photon'' in ''The Photon and Poincare Group'',Valeri V. Dvoeglazov
(Ed.). Nova Science Publishers, NY ,ISBN 1-56072-718-7.

[13]\ See http://www.elec.qmw.ac.uk/staffinfo/mark/mot

\end{document}